\documentclass[]{spie}  %>>> use for US letter paper
\usepackage{amsmath,amsfonts,amssymb}
\usepackage{graphicx}
\usepackage[colorlinks=true, allcolors=blue]{hyperref}
\usepackage{subcaption}
\title{Scientific performances of the XGIS instrument on-board THESEUS}

\author[1,2]{E.~Arrigoni}
\author[1]{S.~Mereghetti}
\author[3,4]{R.~Campana}
\author[3]{C.~Labanti}
\author[5]{A.~Pisapia}
\author[3]{P.~Calabretto}
\author[3]{G.~Mattioli}
\author[3]{E.~Virgilli}
\author[3]{L.~Amati}

\affil[1]{INAF/IASF-Milano, via A. Corti 12, I-20133 Milano, Italy}
\affil[2]{Università degli Studi di Milano-Bicocca, Piazza della Scienza 3, I-20126 Milano, Italy}
\affil[3]{INAF/OAS-Bologna, via P. Gobetti 101, I-40129 Bologna, Italy}
\affil[4]{INFN, Sezione di Bologna, Viale Berti Pichat 6/7, I-40127 Bologna, Italy}
\affil[5]{Università degli Studi di Ferrara, Via G. Saragat 1, I-44122 Ferrara, Italy}

\authorinfo{Further author information: \\ %(Send correspondence 
E-mail: edoardo.arrigoni@inaf.it, sandro.mereghetti@inaf.it}

\begin{document} 
\maketitle

\begin{abstract}
The Transient High Energy Sky and Early Universe Surveyor (THESEUS) satellite, currently undergoing a Phase-A study for the M7 ESA mission selection, will mainly rely on the XGIS (X and Gamma Imaging Spectrometer) instrument to discover and precisely localise gamma-ray bursts and other high-energy transients. The XGIS instrument has been designed to provide accurate sky images in a wide field of view ($\gtrsim$2 steradians)  in the 2–150 keV band, coupled to coverage of nearly half of the sky with good timing and spectroscopic capabilities up to several MeV. This will be achieved thanks to the combination of two identical cameras, based on the coded mask imaging technique, oriented to provide partially overlapping fields of view. We present the capabilities of the XGIS,  focussing in particular on the sensitivity and source location accuracy. We discuss the expected scientific advances in the GRB field, estimated through detailed Monte Carlo simulations which take into account the design of the XGIS imaging system and the properties of the GRB population at high redshifts.
 
\end{abstract}


\section{INTRODUCTION}
\label{sec:intro}  % \label{} allows reference to this section

The \emph{X and Gamma Imaging Spectrometer} (XGIS) is a wide field of view instrument designed to detect and localize, through coded mask imaging, gamma-ray bursts (GRB) and other transients in the 2--150 keV sky, as well as to provide timing and  spectroscopy up to 10 MeV. 
It is one of the three instruments on board the \emph{Transient High Energy Sky and Early Universe Surveyor} (THESEUS) satellite \cite{2018AdSpR..62..191A}, a mission proposed in response to the European Space Agency call for the 7$^{th}$ Medium Size mission of the Cosmic Vision program, with  a launch foreseen in 2037. 
The XGIS will operate in synergy with the two other instruments on board THESEUS: the Soft X-ray Imager (SXI\cite{2021SPIE11444E..2LO}, 0.3--5 keV) providing images on a field of view of $\sim$30$\times$60 deg$^2$, thanks to micropore optics in lobster-eye configuration, and the Infrared Telescope (IRT\cite{2021SPIE11444E..2MG}), a 0.7-m class telescope providing both imaging and spectroscopic ($R\sim400$) capabilities in the 0.7--1.8 $\mu$m range.

THESEUS will have rapid repointing capabilities in order to quickly observe with the IRT the GRBs (and other high-energy transients) discovered by the XGIS and SXI instruments and measure their redshift. 
It will be placed on a low-Earth equatorial orbit with small inclination ($<6^{\circ}$), thus ensuring a low and stable background, and will operate for a nominal mission duration of four years.  
The THESEUS scientific payload and mission profile have been optimized for the detection and characterization of high redshift ($z>6$) GRBs and for the identification of the electromagnetic counterparts of gravitational wave events. The scientific objectives of the THESEUS mission are extensively described in refs.\cite{2018AdSpR..62..191A,2021ExA....52..219T,2021ExA....52..245C,2021ExA....52..309M,2021ExA....52..277G} .

Before presenting the expected imaging capabilities (Section~\ref{sec:fov}), source location accuracy  (Section~\ref{sec:sla}), and sensitivity (Section~\ref{sec:sens}) of the XGIS, we give in Section~\ref{sec:xgis} a brief overview of the instrument design (see ref. \cite{2021SPIE11444E..2KL} for a more detailed description).

\begin{figure}
\centering
\begin{subfigure}{.5\textwidth}
  \centering
  \includegraphics[width=.7\linewidth]{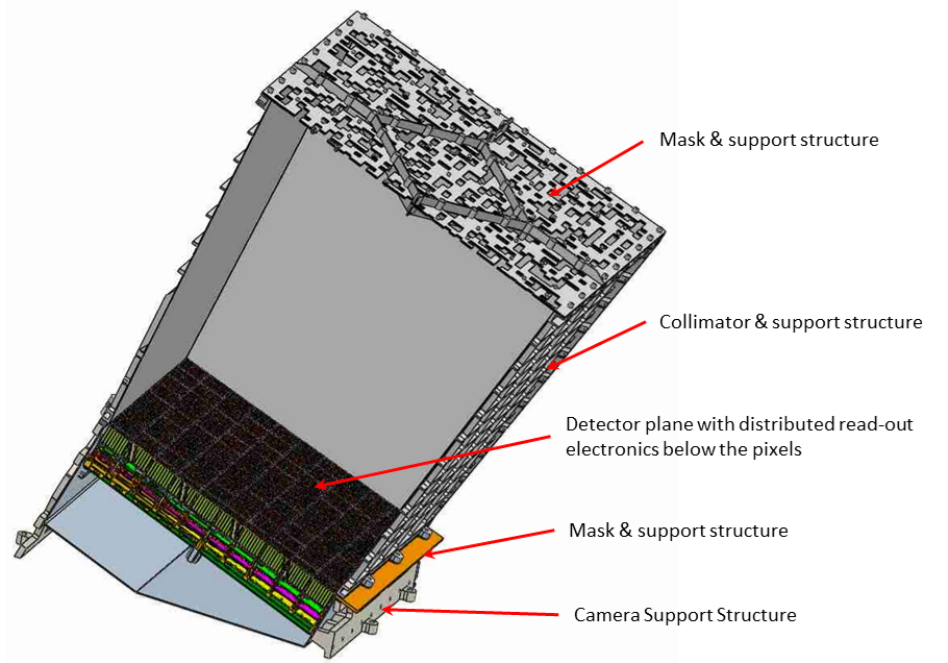}
  \caption{One XGIS camera}
  \label{fig:xgis}
\end{subfigure}%
 \begin{subfigure}{.5\textwidth}
  \centering
  \includegraphics[width=.5\linewidth]{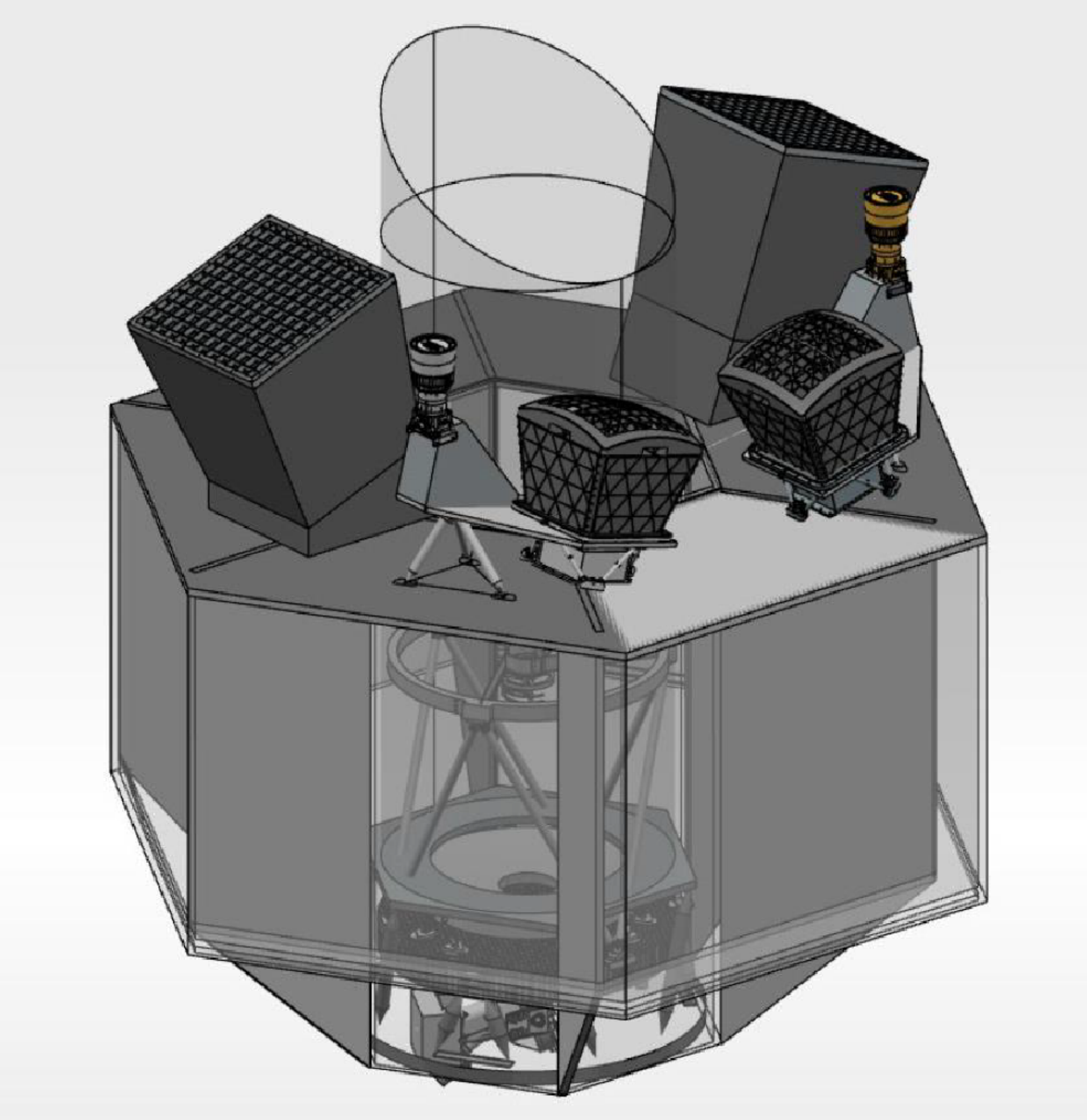}
  \caption{THESEUS satellite}
  \label{fig:theseus}
\end{subfigure}
\caption{Proposed designs for the XGIS cameras and the THESEUS spacecraft}
\label{fig:test}
\end{figure}

\section{The XGIS instrument}
\label{sec:xgis}  % \label{} allows reference to this section

The XGIS instrument is composed of two identical cameras (or ``units", Fig.~\ref{fig:xgis}), two Power Supply Units and one Data Handling Unit (DHU). 
The cameras will provide imaging in the energy range between 2 keV and $\sim$150 keV using the coded mask technique.  The two units will be pointed at directions offset by $\pm20^{\circ}$ from the pointing direction of the IRT  (Fig.~\ref{fig:theseus}).  

%This allows to obtain an  total imaging field of view of $\sim$2 sr by means of the coded mask technique for energies up to $\sim$150 keV. At higher energies (up to several MeV) the XGIS provides only coarse localization capability over a wider FoV ($\sim$4 sr), but still retaining good spectral and timing capabilities.

Each XGIS camera is based on a detection plane made of 80$\times$80 pixels placed at a distance of 63 cm from a coded mask. %composed of square elements of tungsten with a thickness of 1 mm. 
Each pixel consists of a CsI(Tl) scintillator bar, with a height of 30 mm and a square cross section of 4.5$\times$4.5 mm$^2$, sensitive in the nominal 30 keV--10 MeV energy range.
The light produced by gamma-ray photons interacting in the scintillator crystals is collected by two silicon drift detectors (SDD, 5$\times$5 mm$^2$, 450~$\mu$m thickness) placed on the top and bottom of each bar. 
The top SDD (facing the coded mask) acts also as a detector for soft X-ray photons in the nominal 2--30 keV range.

%sensitive in the nominal 30 keV-10 MeV  energy range. The SDD on the top of each bar, besides being  used as a readout for the scintillator light, acts as a detector for soft X-ray photons. Its thickness of 450 $\mu$m provides a high efficiency  in the nominal 2-30 keV  energy range. The modules composing the detection plane have a pitch, defined as the distance between the centers of adjacent pixels, of 5 mm.  The space between modules, required for mechanical and signal read out reasons, has a width equal to the pitch. Thus, from the point of view of the coded mask imaging, the detector array can be considered as a matrix of 89$\times$89 elements (including nine ``dead'' rows and nine ``dead'' columns between the modules) with overall dimensions of 44.5$\times$44.5 cm$^2$.

The coded mask consists of a square array of tungsten elements (1 mm thick) with overall dimensions 
%\cite{gasent20}. The  dimensions of  the mask pattern are 
$\sim$56$\times$56~cm$^2$, supported by a mechanical structure that connects it to the detector and also acts as a passive shield to delimit the field of view at low energies.  The baseline mask aperture is a   pseudo-random self-supporting pattern of 51$\times$51 elements, with open fraction of 41\%.

The DHU, besides all the standard control and data management functions, has the fundamental task of running the on-board software for the GRB trigger and localization using the data of the two XGIS cameras. It will also act as ``master" DHU to combine the information from the three instruments on THESEUS, provide slew requests to the spacecraft, and produce the information on GRB localization to be downloaded in real time.

%The mask pattern design and the dimensions of the mask elements will be optimized during the next phases of the project. 

%The XGIS instrument also comprises two Power Supply Units (one for each camera) and a Data Handling Unit. The latter, 

%The XGIS has been conceived for two complementary aspects: {\it i)} imaging, for source identification and accurate localization and  {\it ii)} spectroscopy, for source characterization over a broad energy range. The first aspect is best achieved in the lower energy range, exploiting the coded mask   in an imaging field of view  defined by the mask and detector geometry,  as described below. On the other hand, in the higher part of the energy range, the absorbing power of the closed  mask elements is too small to significantly modulate the incoming radiation and  the passive shielding, that limits the imaging field of view, becomes increasingly transparent with energy. As a result, the XGIS can detect high-energy radiation also from sources outside the imaging  field of view. For such sources,  only some limited positional information can be derived (e.g., by comparing the count rates in the two cameras), but full timing and spectral information will be obtained. For simplicity, the values  listed in Table 1   refer to the imaging and non-imaging performances  with an energy division at 150 keV,  but it should be kept in mind that this is just a conventional energy threshold and, in reality, there is a gradual transition between the two regimes.

\section{Field of view and angular resolution}
\label{sec:fov}  % \label{} allows reference to this section

The field of view (FoV) of coded mask imaging telescopes depends on the dimensions of mask and detector and on the distance between them. The sensitivity is nearly constant in the central part of the FoV, corresponding to directions for which the whole detection plane collects source photons that passed through the coded mask  (fully coded FoV).  The partially coded FoV corresponds instead to directions for which only part of the detector collects source photons modulated by the mask pattern. In the partially coded FoV the sensitivity gradually decreases with off-axis angle because the detector area used for imaging is reduced. %that collects source photons modulated by the mask aperture. 
%used to record the source counts shrinking, as illustrated in the left panel of Fig.~\ref{fig:fov}. 

\begin{figure}[ht!]
    \centering
    \includegraphics[angle=-90,width=.8\linewidth]{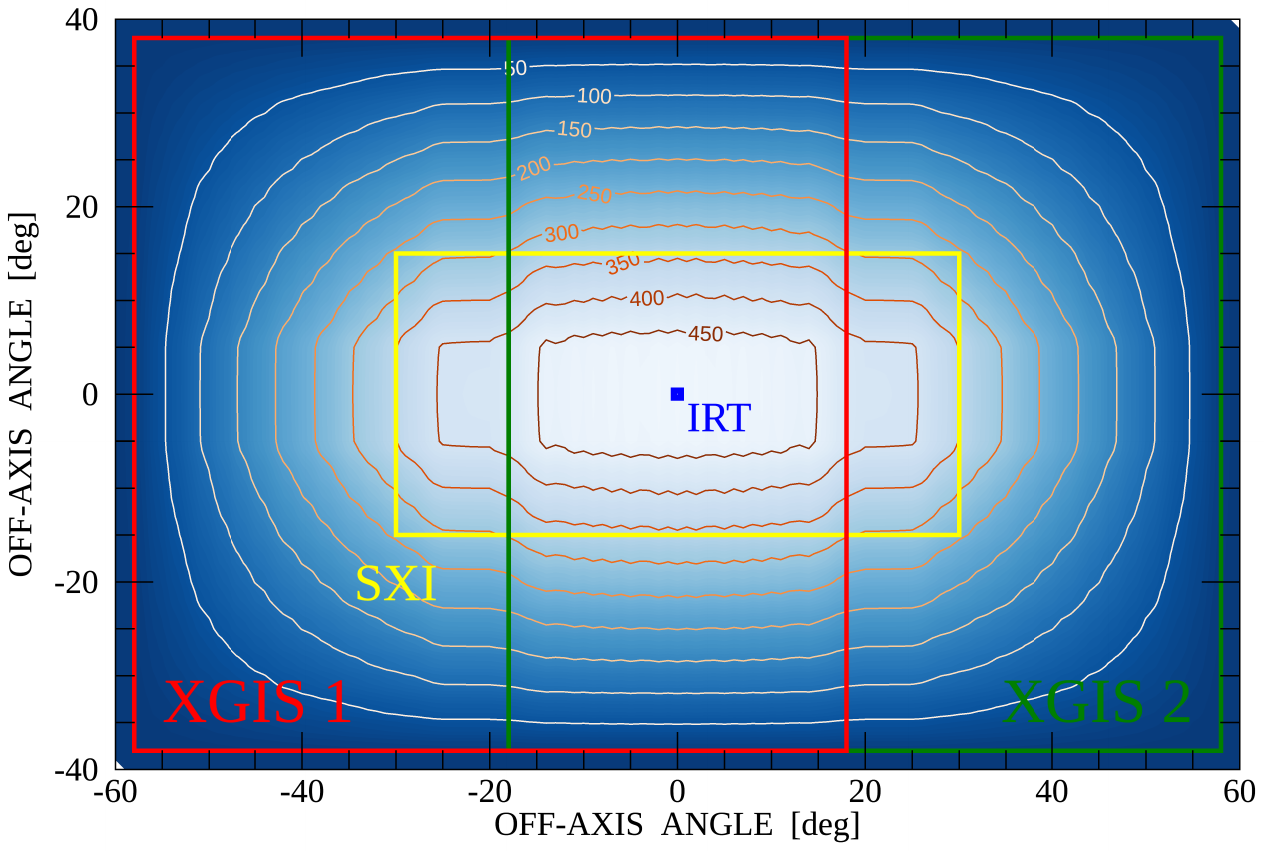}
    \caption{Variation with the off-axis angles of the total effective area (at 10 keV) resulting from the combined use of the two XGIS units. The (partiallly overlapping) fields of view of the  two units  are indicated by the red and green squares.
    %The small dashed squares indicate the  
    %partially coded FoV (77$\times$77 deg$^2$, solid lines) and 
    %fully coded FoV (11$\times$11 deg$^2$). 
    The yellow rectangle is the SXI FoV (61$\times$31 deg$^2$) while the small blue square marks the pointing direction of the IRT.  
    %{\bf   REDONE WITH NEW AREA AND 41\% MASK ** }
    }
    \label{fig:fov}
\end{figure}

\begin{figure}[ht!]
    \centering
    \includegraphics[angle=-90,width=.5\linewidth]{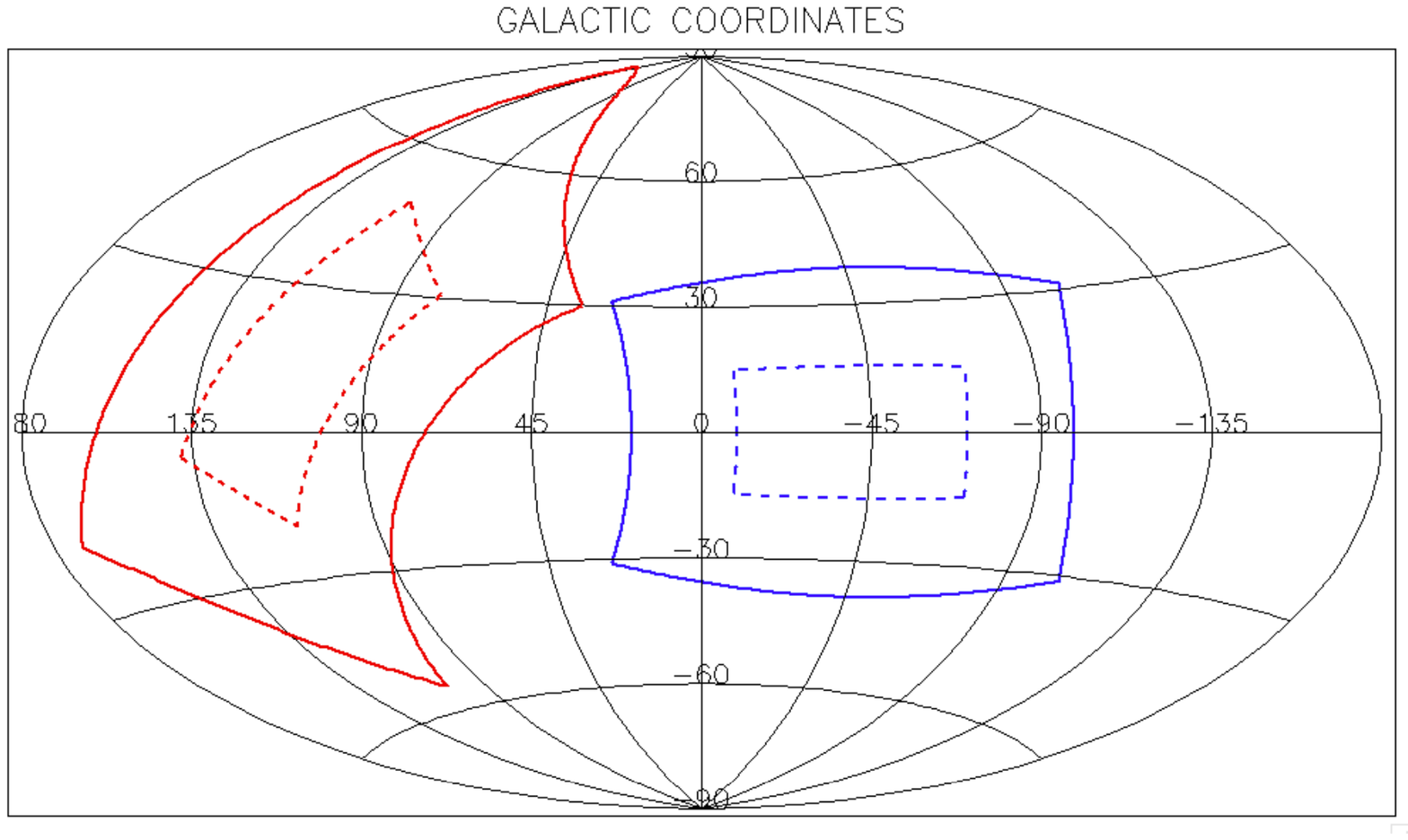}
    \caption{Fields of view of XGIS (solid) and SXI (dashed) projected on a whole  sky map for two representative pointing directions.}
    \label{fig:fovex}
\end{figure}

Each XGIS camera has a fully coded FoV of 
%
%The FCFoV and PCFoV of a single XGIS camera are concentric  squares with sides of  $2\arctan \displaystyle \frac{(D_M-D_D)/2}{H}   
$\sim11^{\circ}\times11^{\circ}$,  
and a partially coded FoV of
%$2 \arctan \displaystyle \frac{(D_M+D_D)/2}{H}       
$\sim77^{\circ}\times77^{\circ}$ (1.6~sr).
Given that the two XGIS cameras point in different directions, offset by $\pm20^{\circ}$, their partially coded FoVs overlap and provide an overall rectangular field of view of  $\sim  77^{\circ}\times117^{\circ}$ = 2.24 sr,  as illustrated in Fig.~\ref{fig:fov}.

The angular resolution $\vartheta_{{\rm res}}$ (not to be confused with the source location accuracy, see Section~\ref{sec:sla}) depends primarily on the size of the mask elements and on the mask-detector distance. Given the mask element size of $\sim$ 11mm,  the XGIS on-axis angular resolution is $\vartheta_{{\rm res}}\sim66'$.

\begin{figure}[h]
    \centering
    \includegraphics[width=1.\linewidth]{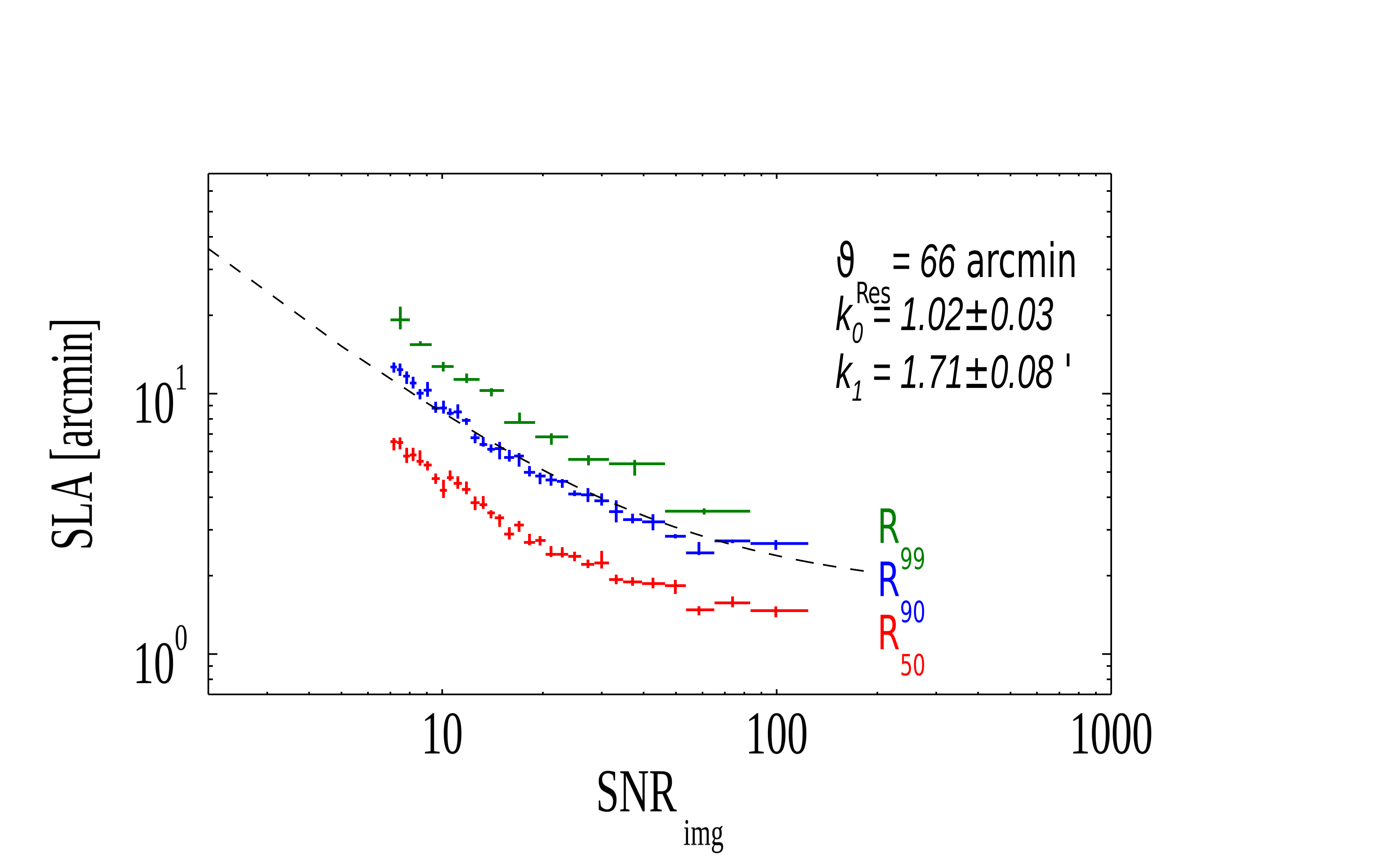} 
    \caption{Source location accuracy of one XGIS camera as a function of the source signal to noise ratio in the reconstructed sky image. The red, blue and green points refer to the 50\%, 90\% and 99\% containment radii, respectively. The dashed line is the best fit to the $R_{90}$ points with equation \ref{eq:sla} and the indicated parameters.}
    \label{fig:sla}
\end{figure}

\section{Source location accuracy}
\label{sec:sla}  % \label{} allows reference to this section

One of the main requirements of the XGIS instrument is to quickly locate GRBs with a positional uncertainty smaller than the field of view of the IRT ($15'\times15'$).  
The source location accuracy (SLA) of a coded mask instrument
%how well it is possible to localise a source does not depend only on the geometric configuration of the mask-detector system (which defines the
is related to the angular resolution, but also depends on the signal to noise ratio (SNR), which quantifies how well the source is distinguished from the background.
From a theoretical point of view, the SLA should be inversely proportional to the SNR of the source\cite{2008ApOpt..47.2739S}, but in practice it is limited by systematic effects.
%the real case both statistical effects intrinsic to the deconvolution process and systematic uncertainties (due to instrumental pointing errors, for instance) limit how well very high SNR sources are localised.
We can therefore describe it with the following  relation:
\begin{equation}
    \mathrm{SLA} = k_0\frac{\vartheta_\mathrm{Res}}{\mathrm{SNR}}+k_1
\label{eq:sla}
\end{equation}

\noindent
where  $\vartheta$ is the angular resolution (see Section \ref{sec:fov}). The parameters $k_0$ and $k_1$  depend on the definition of SLA and SNR and  can be determined through imaging simulations.
In the following we describe the SLA in terms of the 90\% containment radius $R_{90}$ (i.e. the radius of a circular region with a 90\% probability of containing the true position of the source) and we use as signal to noise ratio the source significance $\mathrm{SNR_{img}}$ in the sky image derived from the deconvolution process. 

We performed extensive simulations of the current baseline mask pattern design  for sources extracted from a state-of-the-art synthetic GRB population\cite{2015MNRAS.448.2514G} and uniformly distributed over the whole sky. For each sky image, we computed the localisation error as the difference between the injected and reconstructed sky position. From the distributions of these errors as a function of SNR we derived the results shown in Fig.~\ref{fig:sla}.

Each blue point represents the $R_{90}$ computed over a subsample of images with similar maximum $\mathrm{SNR}_\mathrm{img}$, with horizontal bars defining the width of this group and the vertical ones the 1$\sigma$ $R_{90}$ error obtained applying bootstrapping analysis on each subsample. 
The  $R_{90}$  points are well fitted by equation \ref{eq:sla} with the parameters indicated in the figure. We also show our estimates relative to the $R_{50}$ (in red) and $R_{99}$ (in green) and associated errors, computed as described in the case of $R_{90}$.
    %\begin{comment}
    %The results of extensive simulations using the current baseline mask pattern design ($\vartheta$=66$'$) and a synthetic population of GRBs based on the studies of \cite{??} are shown in Figure \ref{fig:sla}. 
    %to study the imaging performance(see THS-INAF-TN-0014), which allows to obtain the expected relation between $\mathrm{SNR_{img}}$ and localisation accuracy (defined as the radius within which you have 90\% probability of finding the actual position of the source).
    
    %Here we show some other useful metrics (50\% and 99\% containment radius) to better quantify the current expectations, along with the associated errors.
    
    %To obtain these results I simulated 10k images and, considering a trigger threshold of $\mathrm{SNR_{img}}\ge7$, ordered them with increasing $\mathrm{SNR_{img}}$ and binned them in groups of $N_\mathrm{img}$ images each. The horizontal error bars in the plot represent the width of each bin considered.
    
    %The error on the containment radius is obtained by extracting $N_\mathrm{img}$ images randomly and allowing repetitions from each group. For each extraction I computed the relevant containment radii and determined for each group their distributions. The vertical bars show the 1$\sigma$, with the 16 (84) percentile of each distribution as lower (upper) limit.   %\end{comment}

\begin{figure}[ht!]
    \begin{subfigure}[b]{.5\textwidth}
    \centering
    \includegraphics[width=1.\linewidth]{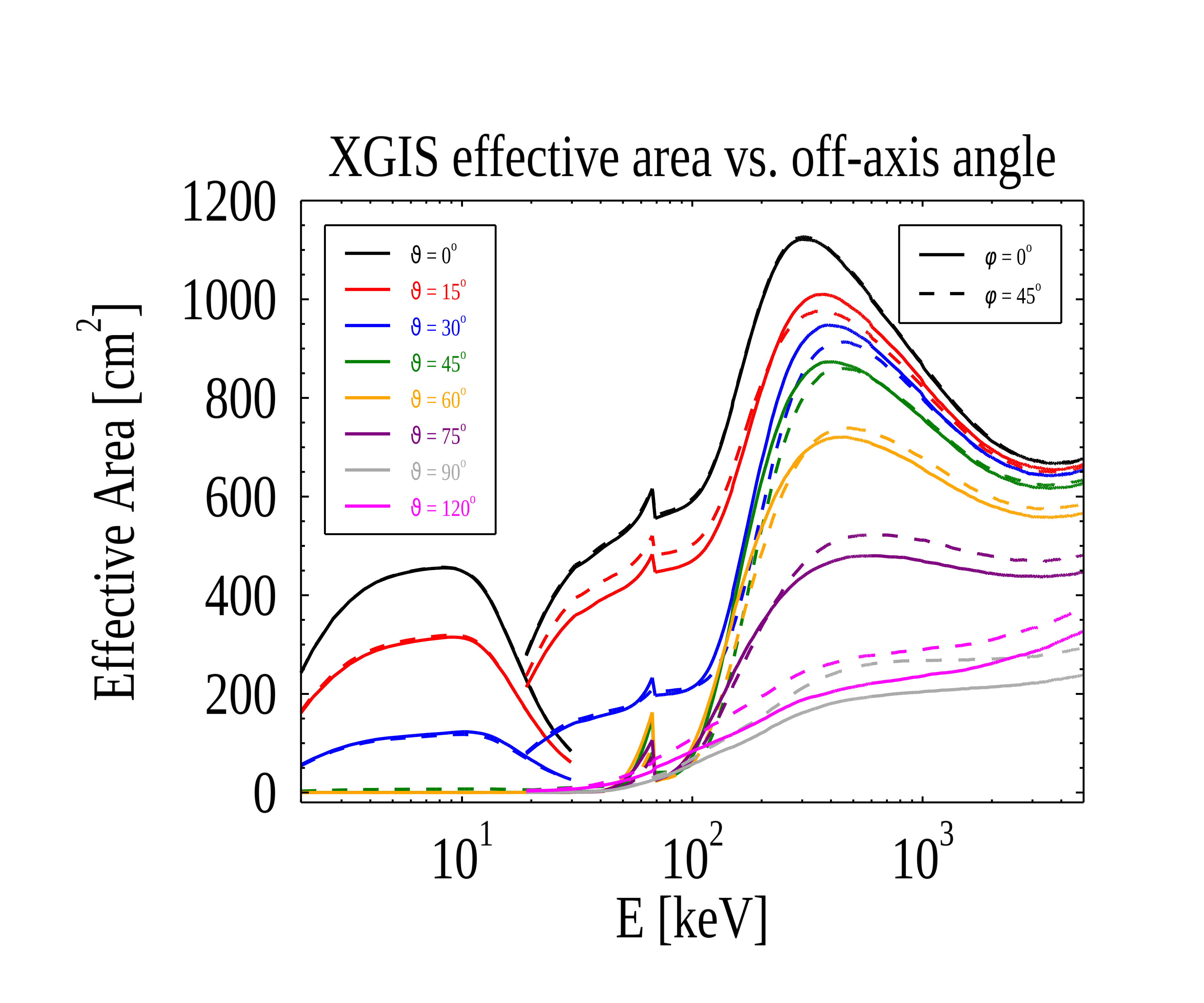}
    \end{subfigure}
    \begin{subfigure}[b]{.5\textwidth}
        \centering
    \includegraphics[width=1.\linewidth]{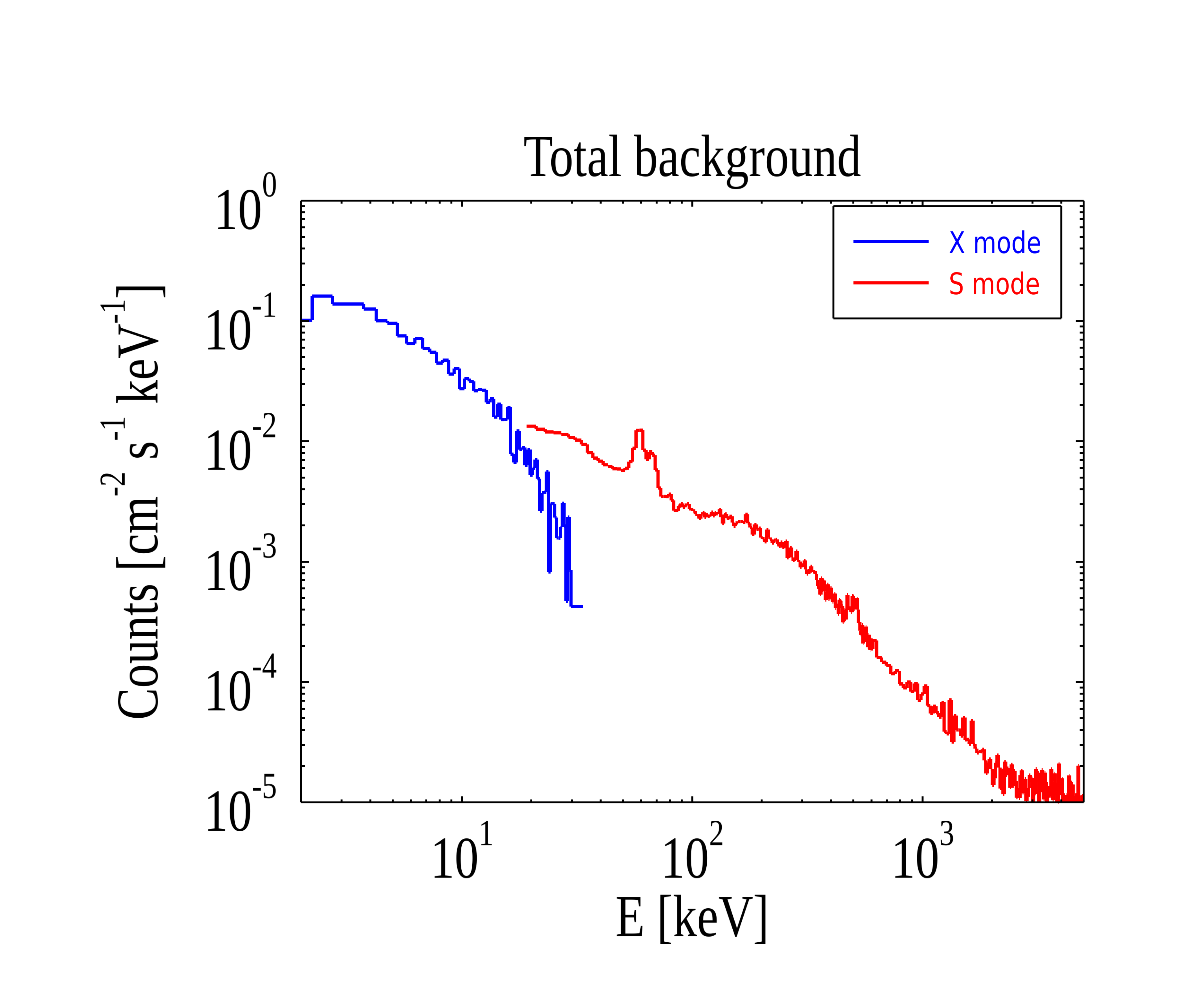}
    \end{subfigure}
    \caption{{\bf Left:} effective area of a single camera as a function of the off-axis position. $\vartheta$ and $\varphi$
    are, respectively, the zenith and azimuth angles with respect to the pointing direction of the camera (i.e. $\vartheta=0^{\circ}$ and $\varphi=0^{\circ}$) is the on-axis direction. $\varphi$=0$^{\circ}$ corresponds to off-axis angles along a side of the detector, $\varphi$=45$^{\circ}$ corresponds to directions along the diagonal.
    {\bf Right:} total expected background in a single camera  for the SDD (X-mode) and the CsI scintillator crystals (S-mode).}
    \label{fig:MCresults}
\end{figure}

\begin{figure}[ht!]
    \centering
    \includegraphics[width=.7\linewidth]{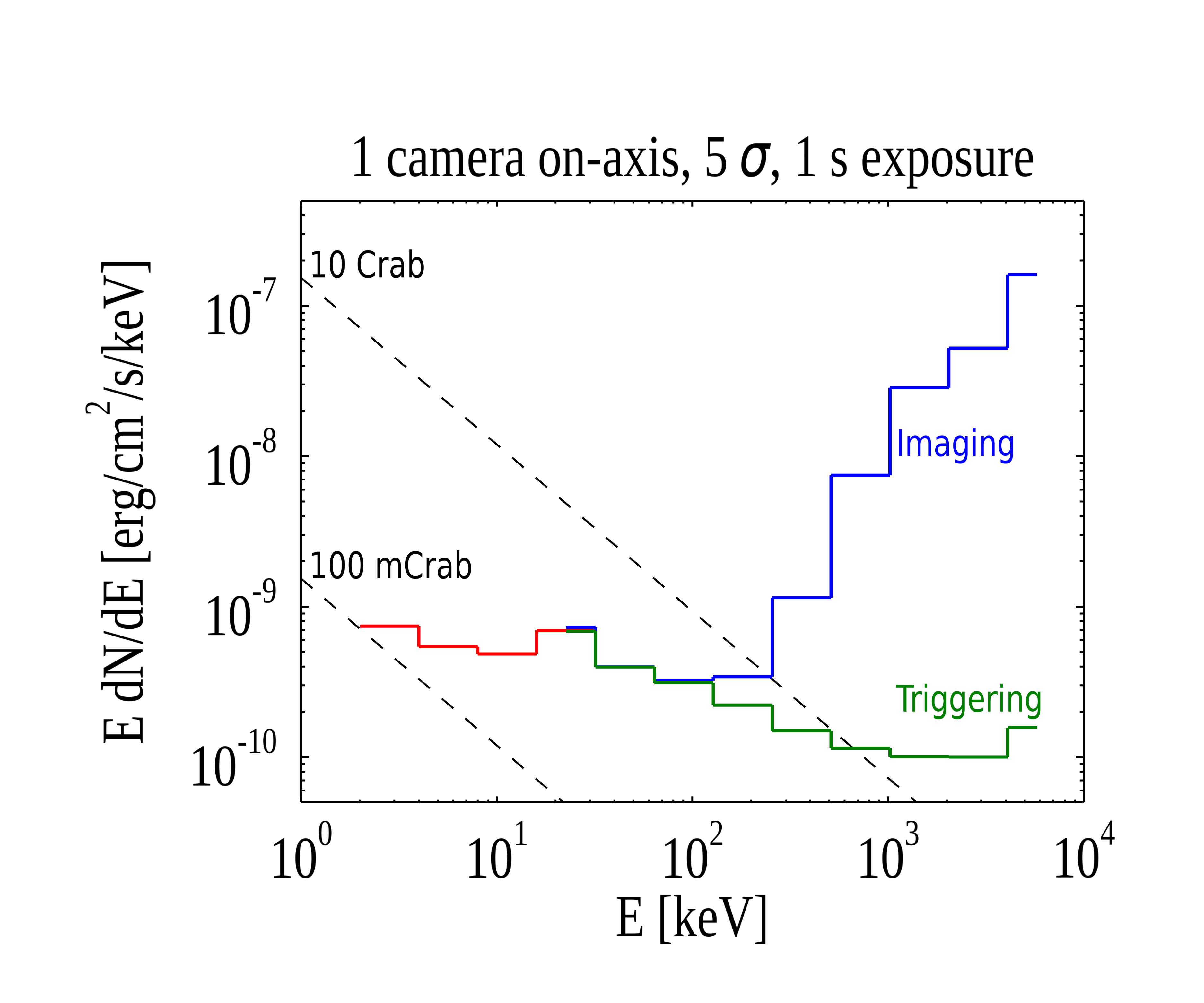}
    \caption{On-axis sensitivity of one XGIS unit for an integration time of 1 s. The red and blue lines are, respectively,  the imaging sensitivities for the SDD detector (X-mode) and  the CsI scintillator (S-mode). The green line is the sensitivity without imaging, at energies where the mask becomes transparent. 
   % For a power law photon index of 2.1. 
    The dashed lines indicate fluxes in Crab units (assuming for the Crab a spectrum with $dN/dE = 9.6 E^{-2.1}$ photons cm$^{-2}$ s$^{-1}$ keV$^{-1}$).}
    %, one Crab unit corresponds to a 2-30 keV flux of .... erg cm$^{-2}$  s$^{-1}$.
%     9.59 E^-2.108   ph/cm2/s/keV         }
    \label{fig:SensHist}
\end{figure}

\begin{table}[ht]
\caption{Scientific performances of the XGIS instrument  } 
\label{tab:sci}
\begin{center}       
\begin{tabular}{ | p{6cm} |p{6cm} |p{4cm} |}
\hline
\rule[-1ex]{0pt}{3.5ex} 
   & Imaging   &  Non-imaging    \\
\hline\rule[-1ex]{0pt}{3.5ex} 
 Energy range & 2--150 keV & $\gtrsim$150 keV    \\
 
\hline\rule[-1ex]{0pt}{3.5ex} 
 Field of view & $11^{\circ}\times11^{\circ}$ fully coded, one camera &  4 sr ($>$20\% efficiency)\\
                     & $77^{\circ}\times77^{\circ}$ total, one camera &    \\
%                     & 80$\times$45 deg$^2$  half  effective area, combined cameras &    \\
                    & $77^{\circ}\times117^{\circ}$  total, combined cameras &    \\

\hline\rule[-1ex]{0pt}{3.5ex} 
Effective area (on-axis, one camera) &   450 cm$^2$ (at 10 keV) & 1100 cm$^2$ (at 300 keV) \\
                    
\hline\rule[-1ex]{0pt}{3.5ex} 
Background (over the whole detector, one camera) &  1400 cts s$^{-1}$ (2-30 keV) &  560 cts s$^{-1}$ (0.15-1 MeV))  \\
            &  700 cts s$^{-1}$ (30-150 keV) &    \\

\hline\rule[-1ex]{0pt}{3.5ex} 
 Sensitivity  (5$\sigma$, 1s, Crab-like spectrum) & $4.3\ 10^{-9}$ erg cm$^{-2}$ s$^{-1}$ (2-30 keV)  & $6.1\ 10^{-8}$ erg cm$^{-2}$ s$^{-1}$  (0.15-1 MeV)     \\
                                               & $2.3\ 10^{-8}$ erg cm$^{-2}$ s$^{-1}$  (30-150 keV) &    \\

\hline\rule[-1ex]{0pt}{3.5ex} 
 Angular resolution  (on-axis, one camera) & $\sim66'$ (FWHM)  &   -  \\
 
\hline\rule[-1ex]{0pt}{3.5ex} 
 Source location  accuracy (90\% c.l. error radius) &   $\sim11'$  (for a source with SNR=7)  &   - \\
         &   $\sim6'$  (for a source with SNR=15)  &   - \\

\hline\rule[-1ex]{0pt}{3.5ex} 
Energy resolution &   $<$ 1200 eV  (FWHM at 6 keV)  &    6\% (FWHM at 500 keV) \\

\hline\rule[-1ex]{0pt}{3.5ex} 
Relative timing accuracy  & \multicolumn{2}{c|}{7 $\mu$s}    \\

\hline\rule[-1ex]{0pt}{3.5ex} 
GRB expected rate (not corrected for duty cycle of observations) &  501 yr$^{-1}$  (long)  &  \\
  &  19 yr$^{-1}$  (long, z$>$6) &   \\
  &  20 yr$^{-1}$  (short) &   \\
\hline

\end{tabular}
\end{center}
\end{table}

\section{Sensitivity}
\label{sec:sens}  % \label{} allows reference to this section
Our sensitivity estimates are based on the effective area $A_\mathrm{eff}$ derived,  %and on the expected background from \cite{BKG}.  $A_\mathrm{eff}$ has been derived 
as a function of energy and for different directions,  with the Monte Carlo simulations 
described in ref.\cite{2021SPIE11444E..8PC} . Besides accounting for detector efficiency, it includes the effect of the coded mask, properly taking into account the aperture pattern, the open fraction $\tau$,  and the transmission  $\eta_\mathrm{O}(E)$ and $\eta_\mathrm{C}(E)$ of the open and closed elements.
The effective area as a function of energy for a single unit is shown in the left panel of Fig.~\ref{fig:MCresults} for different off-axis angles. The two sets of curves refer to detections in the SDD (X-mode, 2--30 keV) and in the CsI scintillator bars (S-mode, 0.03--10 MeV). The great increase in effective area above $\sim$200 keV, is caused by the fact that the mask and its supporting structure become transparent at these energies. Therefore, the source photons are collected by the whole detector, but without being modulated by the mask pattern. Although images cannot be done, this energy range is still useful for spectral and timing studies and contributes to the trigger sensitivity for GRBs.

For the background $B(E)$ (right panel of Fig.~\ref{fig:MCresults})
%measured in [counts cm$^{-2}$ s$^{-1}$ keV$^{-1}$], 
we use the model derived in Refs.\cite{2021SPIE11444E..8PC,2022hxga.book...39C}. This model includes both the photons from the cosmic diffuse X-ray  background, which is dominant at low energies, and the contribution of cosmic ray particles, which becomes significant above $\sim$100 keV.

The performance of a coded mask imaging instrument depends on the contrast between the part of detector that is directly illuminated by a source flux passing through the open mask elements and the part of detector shadowed by the closed mask elements.  In an ideal mask, the open elements are completely transparent to incoming radiation and the closed ones are completely opaque. In the case of the XGIS, this is a good approximation at low energies ($\lesssim$150 keV), but, since the tungsten mask thickness is only 1 mm, it becomes increasingly transparent at higher energies. Therefore, the source photons that pass through the closed mask elements contribute to the background in the deconvolved images. We have properly taken into account this effect to compute the imaging sensitivity, as explained in  Appendix \ref{sec:app_sens}.

\begin{figure}[ht!]
    %\begin{subfigure}[b]{.5\textwidth}
    %\centering
    %\includegraphics[width=1.\linewidth]{fig/XGIS_offa_photsens_v2-1comp.png}
    %\end{subfigure}
    %\begin{subfigure}[b]{.5\textwidth}
    \centering
    \includegraphics[width=1.\linewidth]{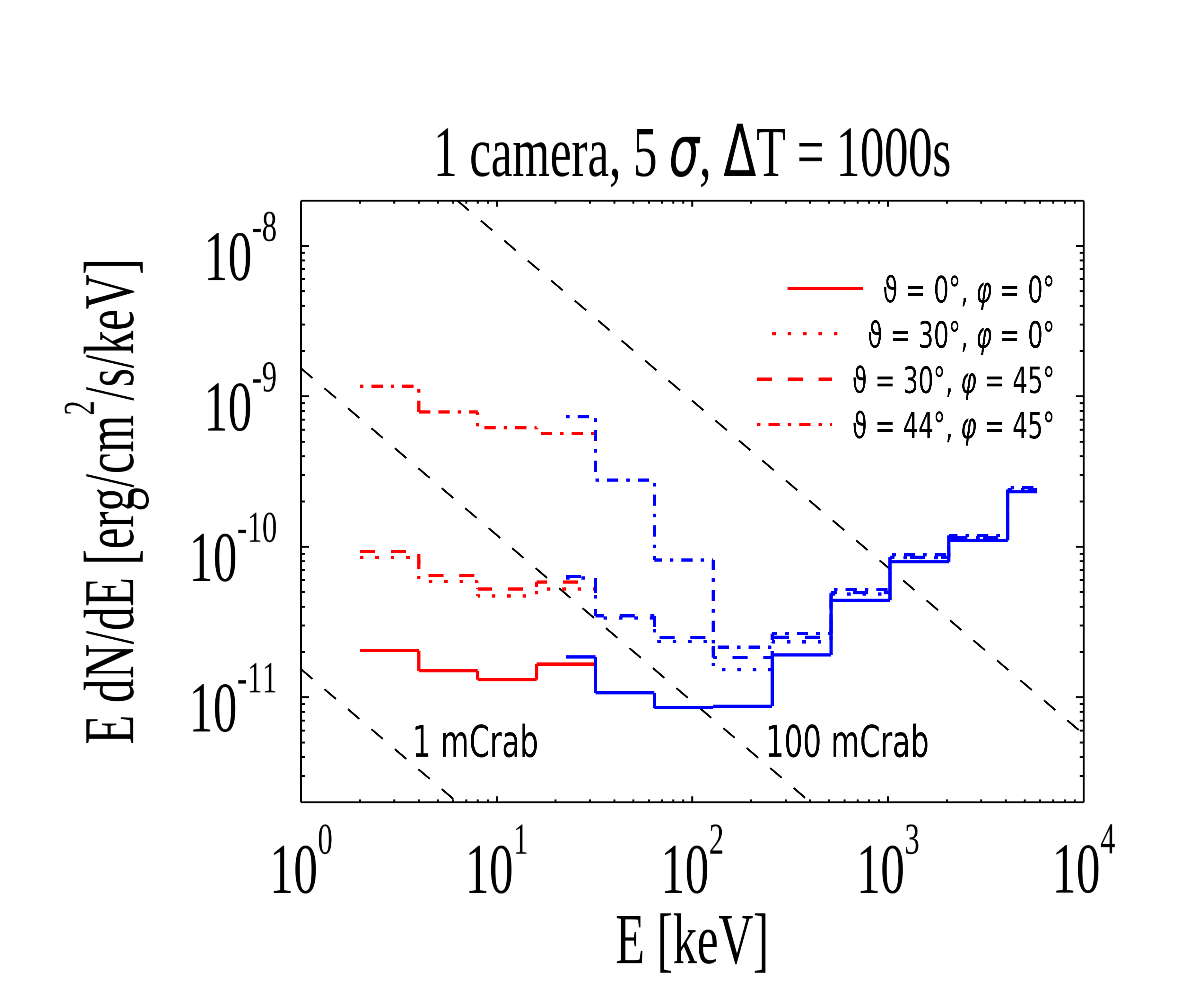}
    %\end{subfigure}
    \caption{Imaging sensitivity for 1 ks exposure   for different positions in the FoV for the SDD (X-mode, red) and the CsI scintillator (S-mode, blue).  $\varphi$=0$^{\circ}$ corresponds to off-axis angles along a side of the detector, $\varphi$=45$^{\circ}$ corresponds to directions along the diagonal.}
    \label{fig:Sens1k_offa}
\end{figure}

The on-axis sensitivity of one XGIS camera for an integration time of 1 s
%(eq. (\ref{eq:ImagSens}) and (\ref{eq:NonImagSens})) 
is shown in Fig.~\ref{fig:SensHist}, where the widths of the histogram bins represent the energy bands over which we integrated detected photons.  The sensitivity for an integration time of 1000 s at different off-axis positions is shown in Fig.~ \ref{fig:Sens1k_offa}. 
These sensitivities (Eqs. \ref{eq:ImagSens} and \ref{eq:NonImagSens}) were computed for  a source with power-law spectrum
%$dN(E)/dE=S$ 
of photon index 2.1.
 
%To apply our result to a variety of sources which will be observed by XGIS, we have also studied how it will perform in different positions in its FoV (fig. \ref{fig:Sens1k_offa}) and for different spectral shapes (see fig. \ref{fig:SensEpeak}).

\begin{figure}[ht!]
    \centering
    \includegraphics[width=.7\linewidth]{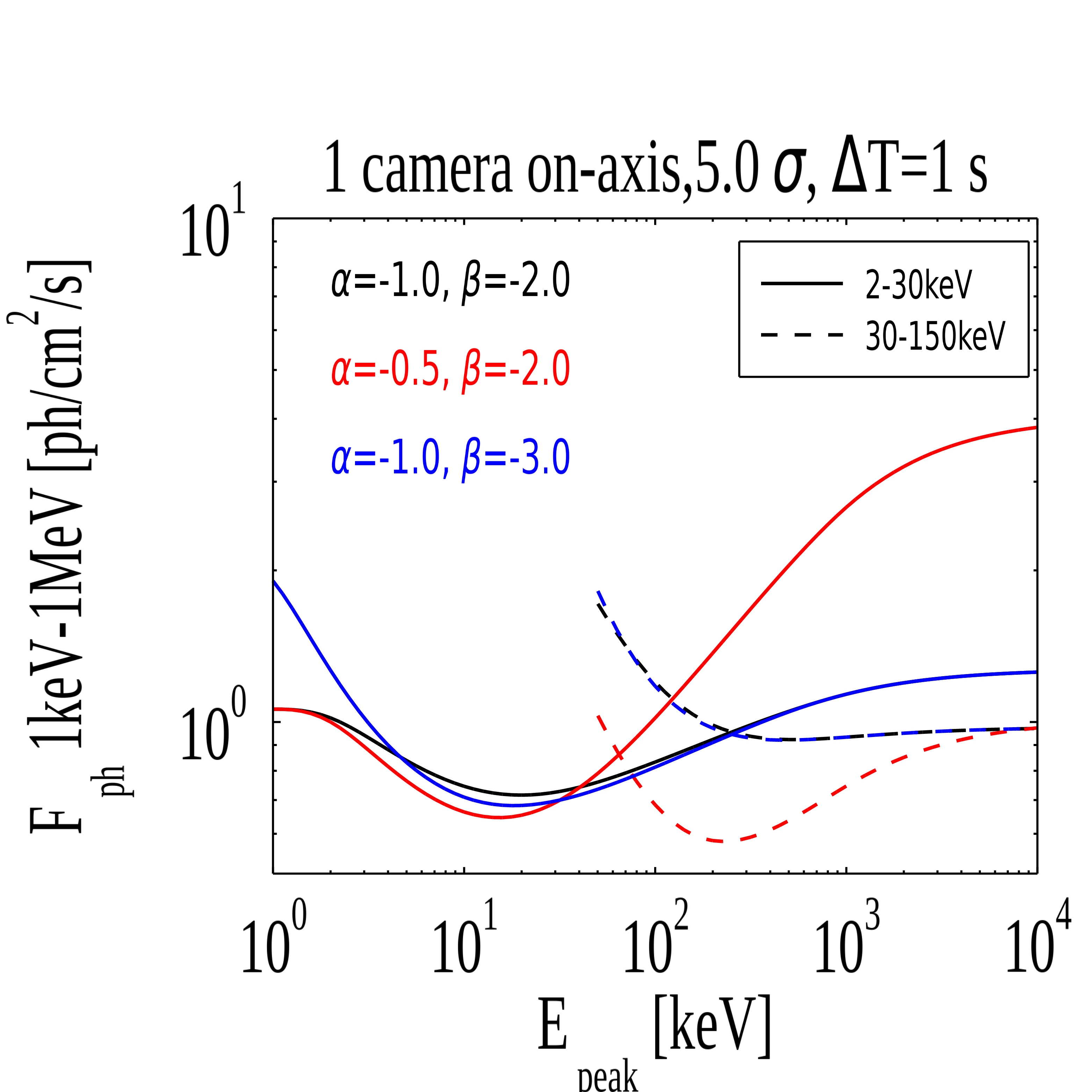}
    %\begin{subfigure}[b]{.5\textwidth}
    %\centering
    %\includegraphics[width=1.\linewidth]{fig/YB_logS-logEpeak_v2-1.png}
    %\end{subfigure}
    %\begin{subfigure}[b]{.5\textwidth}
    %    \centering
    %\includegraphics[width=1.\linewidth]{fig/YB_logS-logEpeak_v2-1b.png}
    %\end{subfigure}
    \caption{Imaging sensitivity in X mode (solid lines) and S mode (dashed lines) as a function of $E_\mathrm{peak}$, $\alpha$ and $\beta$ parameters in a Band spectrum. S mode becomes more sensitive only for GRB with greater $E_\mathrm{peak}$ and $\alpha$.}
    \label{fig:SensEpeak}
\end{figure}

The spectra of GRBs are usually described in terms of two smoothly connected power laws with photon index $\alpha$ and $\beta$ at low and high energy, respectively, with $E_{peak}$ representing the energy at which the energy spectrum peaks \cite{1993ApJ...413..281B}. We show in Fig.~\ref{fig:SensEpeak} how the XGIS sensitivity varies as a function of the source spectral parameters. The low energy range covered in the X-mode gives an excellent sensitivity to sources with soft spectra, favoring the detection of high redshift GRBs, which is one of the key objectives of the THESEUS mission.\\

%FARE COMBINED SENSITIVITY OF TWO UNITS....

\section{Conclusions}
\label{sec:conc}  

The performance of the XGIS, summarized in Table~\ref{tab:sci}, makes it an ideal instrument to fulfill the scientific objectives of the THESEUS mission. In particular, the wide field of view, broad energy range, high sensitivity,  and good source location accuracy will allow the collection of a large sample of high redshift GRBs and the detection of electromagnetic counterparts of gravitational waves and neutrino events provided by the next generation facilities.

\begin{figure}[ht!]
    \centering
    \includegraphics[width=.7\linewidth]{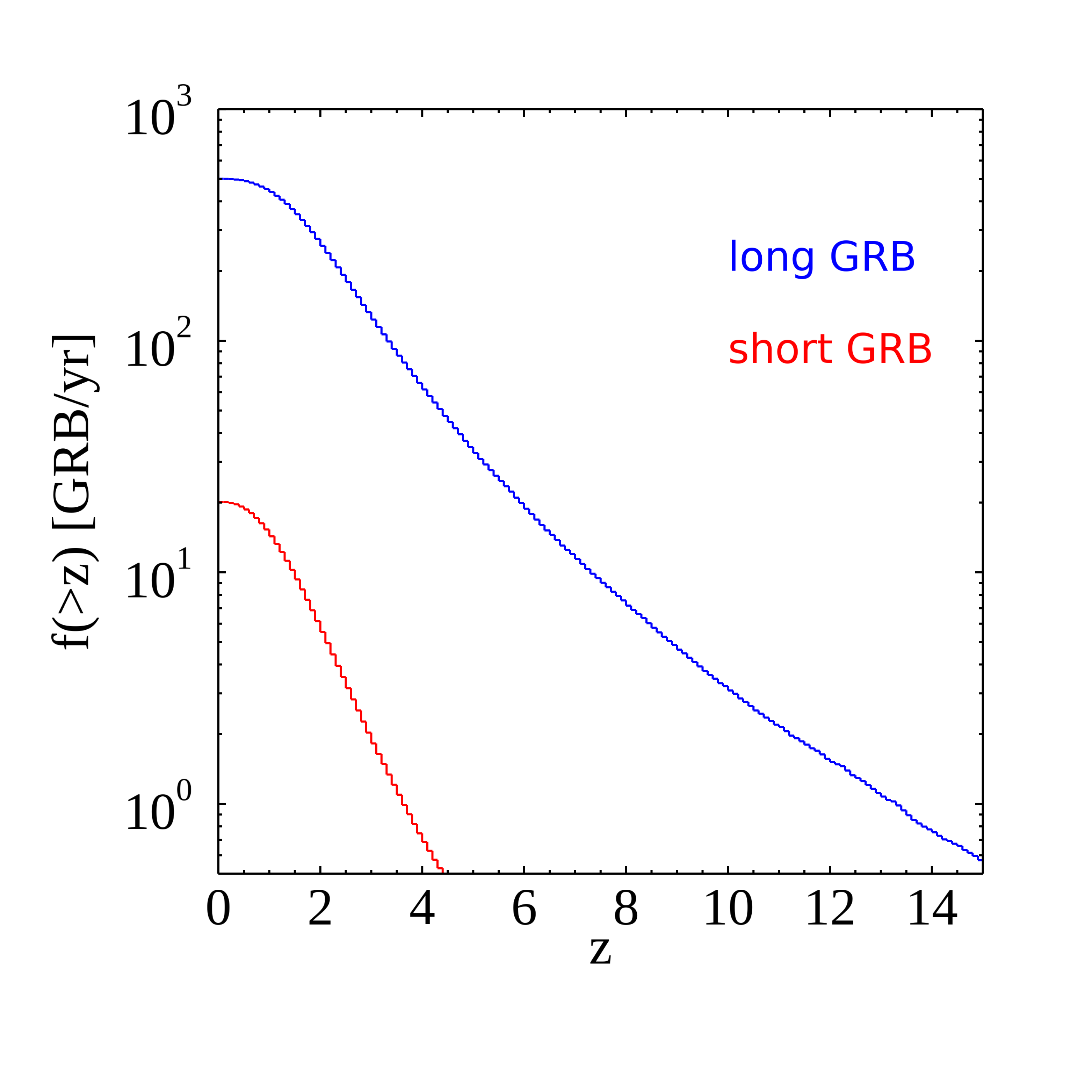}
    \caption{Cumulative distribution for long (blue) and short (red) GRB detected using a combined SNR threshold (fixed at 7) and assuming 100\% duty cycle.}
    \label{fig:redshiftDistr}
\end{figure}

Based on the best knowledge currently available on the populations of long and short GRBs \cite{2015MNRAS.448.2514G,2016A&A...594A..84G}, we have estimated with detailed simulations, the expected rates of detections of these events. In these simulations, we have taken into account the dependence of effective area as a function of energy and position in the field of view, as well as the expected distribution of spectral shapes and durations of GRBs as a function of their redshift. To detect a GRB we adopted a  threshold signal to noise ratio $n$ of 7 (we used $n=(n_1^2 + n_2^2)^{1/2}$, to combine the signal to noise ratios of the two units in the overlapping FOV region). 
The resulting rates of GRBs detected and localised by the whole XGIS instrument (2 units combined) is of 500 long GRBs per year (of which 19 above redshift $z=6$) and 20 short GRBs per year. Their redshift distributions are plotted in  Fig.\ref{fig:redshiftDistr}. Of these, around 70\% are observed with a SNR which corresponds, through eq.\ref{eq:sla} with parameters determined in section \ref{sec:sla}, to an $R_{90} <7'$ and will therefore very likely lead to a successful follow up observation with the on-board IRT.
For simplicity, the rates given here are based on an  observing efficiency $\epsilon_{\rm obs}$ = 100\% (i.e., they are not corrected for time intervals in which the instrument is not operating or its field of view is partially obstructed). These effects  have been evaluated through a Mission Observation Simulator, developed by ESA,  which properly took into account the  orbital and viewing constraints, the pointing and observation strategy, the times required for slews, and  all the other relevant operational aspects of the THESEUS mission. The results indicate that the final rates are adequate to fulfill with a good margin the scientific requirements of the THESEUS mission, providing an unprecedented sample of high redshift GRBs and electromagnetic counterparts of gravitational waves events.

%the two XGIS cameras on THESEUS will be able to detect and localise more than 700 long GRBs per year, 35 of which are expected to have a redshift greater than 6 (which corresponds to the Universe less than a billion years after the Big Bang), and around 30 short GRBs/yr. This rates assume a 100\% duty cycle of the satellite.%, and any loss of efficiency due to the observation dead time is taken into account by the Mission Operation Simulator (MOS). From past and current experience with coded mask instruments we can estimate a realistic duty cycle of around ??\%. %RATE DI LUNGHI E CORTI IN FUNZIONE DEL REDSHIFT   (TRIGGER SENS. 2 CAMERE, 100\% DUTY CYCLE) 

%\begin{table}[]
%    \centering
%    \begin{tabular}{lc}
%\hline
%Instrument    & XGIS \\
%\hline
%Energy range         & 2 keV - 8 MeV \\
%Time resolution & 500 ns ??\\
%Field of view & \\
%\hline
%A$_\mathrm{eff}$ @ 10 keV         &  450 cm$^2$\\
%A$_\mathrm{eff}$ @ 300 keV         &  1100 cm$^2$\\
%Bkg in 2-30 keV & 1350 cts/s\\
%Sensitivity in 2-30 keV & ---\\
%\hline
%Long GRBs & 730 per year\\
% $z > 6$& 35 per year\\
%Short GRBs & 29 per year\\
%R$_{90}$ @ SNR=7 & 10'\\
%    \end{tabular}
%    \caption{  }
%    \label{tab:XGISspecs}
%\end{table}

%\textbf{Compare with real images for sources of flux Smin to check finite size element effects?}

\appendix    %>>>> this command starts appendixes

\section{Derivation of sensitivity}
\label{sec:app_sens}

%\subsection{Imaging}
%\label{subsec:ima}  % \label{} allows reference to this section

The imaging sensitivity of a coded mask instrument depends on the contrast between the  part of detector illuminated by the source through the open mask elements and  that shadowed by the closed elements, which collect $N_O$ and $N_C$ counts, respectively.
For a source of specific flux $dN(E)/dE=F_S$[photons cm$^{-2}$ s$^{-1}$ keV$^{-1}$] observed for a time interval $\Delta t$ in the energy range $\Delta E$ with a background $F_B(E)$[counts cm$^{-2}$ s$^{-1}$ keV$^{-1}$], these can be written as:

\begin{equation}
    \begin{array}{@{}l@{}}
    N_O=\int  [A_\mathrm{eff}^\mathrm{open}(E)F_S(E)+\tau A_\mathrm{geom}F_B(E)]d\mathrm{t} d\mathrm{E}\\\    
    N_C=\int  [A_\mathrm{eff}^\mathrm{closed}(E)F_S(E)+(1-\tau) A_\mathrm{geom}F_B(E)] d\mathrm{t} d\mathrm{E}\
    \end{array}\,
    \label{eq:Counts_OC}
\end{equation}

\noindent
where $\tau$ is the mask open fraction (i.e. the ratio between the open and total mask surface),  $A_\mathrm{geom}$ is the detector geometric area, 
and we have split the total effective area $A_\mathrm{eff}=(\eta_O\tau+\eta_C(1-\tau))\varepsilon A_\mathrm{geom}$ into two components referring respectively to the illuminated and shadowed portions of the detector. $\eta_O$ and $\eta_C$ are the transparency of the open and closed mask elements (i.e. an ideal mask has $\eta_O=1$ and $\eta_C=0$,  independent of energy) and $\varepsilon$ is the detector efficiency. For simplicity, we do not indicate that these quantities depend also on the direction in the field of view.

Starting from the total effective area $A_\mathrm{eff}$ obtained from Monte Carlo simulations, we estimate its two components related to the open and closed parts of the detector as:

 \begin{equation}
    \begin{array}{@{}l@{}}
    A_\mathrm{eff}^\mathrm{open}(E)= A_\mathrm{eff}(E)\displaystyle\frac{\tau\eta_O(E)}{\tau\eta_O(E)+(1-\tau)\eta_C(E)}\\
     A_\mathrm{eff}^\mathrm{closed}(E)= A_\mathrm{eff}(E)\displaystyle\frac{(1-\tau)\eta_C(E)}{\tau\eta_O(E)+(1-\tau)\eta_C(E)}
    \end{array}\,
\end{equation}

%We define the transmission coefficients $\bar\eta_{O/C}$ in the energy interval $\Delta E$, weighted with the source spectrum $S(E)$ and the detector response $\varepsilon_\mathrm{det}$, as follows
%\noindent

Defining the average effective areas in the range $\Delta E$ and weighted with the source spectrum as

 \begin{equation}
    \begin{array}{@{}l@{}}
    %\bar{A} = \displaystyle \frac{\int A_\mathrm{eff}(E)S(E) d\mathrm{E}}{\int S(E) d\mathrm{E}}\\
    %\bar{A} = \displaystyle \frac{\int A_\mathrm{geom}\epsilon (E)S(E) d\mathrm{E}}{\int S(E) d\mathrm{E}}\\
    %\overline{\varepsilon\eta}_{O/C} = \displaystyle \frac{\int \eta_{O/C}(E)~\varepsilon (E)S(E) d\mathrm{E}}{\int S(E) d\mathrm{E}}\\
    \overline{A}_\mathrm{eff}^\mathrm{open/closed} = \displaystyle \frac{\int {A}_\mathrm{eff}^\mathrm{open/closed} (E)S(E) d\mathrm{E}}{\int S(E) d\mathrm{E}}\\
    \end{array}\, ,
\end{equation}

\noindent
%and in a similar way $\bar{A}^\mathrm{open}_\mathrm{eff}$, $\bar{A}^\mathrm{closed}_\mathrm{eff}$, $\bar{\eta}_O$, and $\bar{\eta}_C$. 
%from which follow the definitions of average effective areas as
%\begin{equation}
%    \begin{array}{@{}l@{}}
%    \bar{A}\mathrm{_{eff}^{open}} = \tau %~A_\mathrm{geom}\overline{\varepsilon\eta}_O\\
%    \bar{A}\mathrm{_{eff}^{closed}} = (1-\tau) ~A_\mathrm{geom}\overline{\varepsilon\eta}_C\\
%    \bar{A}\mathrm{_{eff}} = \bar{A}\mathrm{_{eff}^{open}}+\bar{A}\mathrm{_{eff}^{closed}}\\
%    \end{array}\,
%    \label{eq:AreaEff}
%\end{equation}
equation~\ref{eq:Counts_OC} can be rewritten as

\begin{equation}
    \begin{array}{@{}l@{}}
    N_O=  [\bar{A}^\mathrm{open}_\mathrm{eff} {S} + \tau A_\mathrm{geom}{B}]  \\\
    N_C= [\bar{A}^\mathrm{closed}_\mathrm{eff} {S} + (1-\tau) A_\mathrm{geom}{B}]  \\\
    \end{array}\,
    \label{eq:Counts_OC2}
\end{equation}

\noindent
where ${S}$ and ${B}$ are the source and background fluences integrated over the interval $\Delta E\Delta t$. 
%The flux $\bar{S}$ and its error $\sigma_{\bar{S}}$
%, and the signal to noise ratio $n$=$\bar{S}$/$\sigma_{\bar{S}}$
Solving equations~\ref{eq:Counts_OC2}, we also obtain the source fluence uncertainty $\sigma_{S}$, assuming Poissonian statistics on measured counts $N_O$ and $N_C$, as a function of $S$:
%The source fluence $S$ and its error $\sigma_{S}$ can be derived from the measured values of $N_O$ and $N_C$ by solving equations~\ref{eq:Counts_OC2}:

\begin{equation}
    \begin{array}{@{}l@{}}
         \bar{S} = \mathrm{K_{T}} (N_O(1-\tau)-N_C\tau) \\
         \sigma_{\bar S} = \mathrm{K_{T}}\sqrt{\mathrm{K_{\tilde S}}\bar S+\mathrm{K_B}} 
         \end{array}\, 
\end{equation}

\noindent
where we have defined the following quantities:
\begin{equation}
    \begin{array}{@{}l@{}}
         %\mathrm{K_{S} } :=  A_\mathrm{geom}~\tau(1-\tau)~(\overline{\varepsilon\eta}_O(1-\tau)+\overline{\varepsilon\eta}_C\tau)~\Delta\mathrm{E} \Delta\mathrm{t}  \\
         %\mathrm{K_B} :=A_\mathrm{geom}~\tau(1-\tau)\bar{B}~\Delta\mathrm{E} \Delta\mathrm{t} \\
         %\mathrm{K_{T}} :=\frac{1}{A_\mathrm{geom}~\tau(1-\tau)~(\overline{\varepsilon\eta}_O-\overline{\varepsilon\eta}_C)~\Delta\mathrm{E} \Delta\mathrm{t}} 

         \mathrm{K_{S} } :=  (\bar{A}^\mathrm{open}_\mathrm{eff}(1-\tau)^2+\bar{A}^\mathrm{closed}_\mathrm{eff}\tau^2) \\
         \mathrm{K_B} :=A_\mathrm{geom}~\tau(1-\tau){B}\\
         \mathrm{K_{T}} := \displaystyle \frac{1}{(\bar{A}^\mathrm{open}_\mathrm{eff}(1-\tau)-\bar{A}^\mathrm{closed}_\mathrm{eff}\tau)} 
    \end{array}\, 
\end{equation}
%which allow us to write the estimated error $\sigma_{\overline{S}}$, assuming Poissonian statistics
%\begin{equation}
%    \begin{array}{@{}l@{}}
%         \bar{S} = \mathrm{K_{T}} (N_O(1-\tau)-N_C\tau) \\
%         \sigma_{\bar S} = \mathrm{K_{T}}\sqrt{\mathrm{K_{\tilde S}}\bar S+\mathrm{K_B}} 
%         \end{array}\, 
%\end{equation}
\noindent
The imaging sensitivity, for a given  signal-to-noise ratio  $n= {\overline S} / \sigma_{\overline S}$, is then given by:

%\begin{equation}
%    n=\frac{\overline S}{\sigma_{\overline S}}=\frac{\overline S}{\mathrm{K_T}\sqrt{\mathrm{K_S}\overline S+\mathrm{K_B}}}
    %=\bar{S}~(\overline{\varepsilon\eta}_O-\overline{\varepsilon\eta}_C)\sqrt{\frac{\tau(1-\tau)~A_\mathrm{geom}~\Delta\mathrm{E} \Delta\mathrm{t}}{(\overline{\varepsilon\eta}_O(1-\tau)+\overline{\varepsilon\eta}_C\tau)~\bar S+\bar B}}.
%    \label{eq:significance}
%\end{equation}

%Assuming a significance threshold $n_\mathrm{thr}$ for a valid detection we can solve the previous equation for $ \overline{S}$ an find
\begin{equation}
    \begin{array}{@{}l@{}}
  {S}_\mathrm{min}= \displaystyle \frac{(n\mathrm{K_{T}})^2\mathrm{K_{S}}+\sqrt{n^4\mathrm{K_{T}}^4 \mathrm{K_{S}}^2+4n^2\mathrm{K_{T}}^2 \mathrm{K_B}}}{2} .\\
   % = \frac{n^2_\mathrm{thr}(\bar\eta_O/\tau+\bar\eta_C/(1-\tau))+\sqrt{(n^2_\mathrm{thr}(\bar\eta_O/\tau+\bar\eta_C/(1-\tau)))^2+4 n_\mathrm{thr}^2(1/\tau+1/(1-\tau))(\bar\eta_O-\bar\eta_C)^2~\bar{B}~A_\mathrm{geom}\Delta\mathrm{E} \Delta\mathrm{t}}}{2(\bar\eta_O-\bar\eta_C)^2~A_\mathrm{geom}~\Delta\mathrm{E} \Delta\mathrm{t}}.
    \end{array}
    \label{eq:ImagSens}
\end{equation}

%where $\overline{S}_\mathrm{min}$ represents the minimum flux (in ph cm$^{-2}$ s$^{-1}$ keV$^{-1}$), averaged in the energy and time interval considered, required to produce an image with a peak of signal-to-noise $n$.

%\subsection{Non-imaging}
%\label{subsec:noima}  % \label{} allows reference to this section

As energy increases,  the  closed mask elements become progressively transparent, and when ${\varepsilon\eta}_C={\varepsilon\eta}_O$ 
%From equation \ref{eq:significance} we can immediately see that, for a given $\overline{S}$ and $\varepsilon_\mathrm{det}(E)$, higher values of $\overline{\eta}_C$ ($\overline{\varepsilon\eta}_C$) produce lower significance peak in the resulting image and, in the limit of $\overline{\varepsilon\eta}_C=\overline{\varepsilon\eta}_O$ (condition equivalent to $\overline{\eta}_C=\overline{\eta}_O$ where $\varepsilon_\mathrm{det}(E)\ne 0$), 
the instrument loses its imaging capability. In the case of the XGIS, this happens for  photons with energy $>$ 200 keV, for which the 1mm thick mask is completely transparent. Still, the detector itself is capable of detecting these photons and the XGIS can operate as a collimated, non-imaging instrument to
%. , as shown in fig. \ref{fig:MCresults}, and can therefore be used to 
study source light curves and spectra.

We can define the non-imaging sensitivity   as the minimum fluence ${S}_\mathrm{NI}$ (integrated over a chosen time and energy interval) that produces an increase in the number of detected counts with significance $n_\mathrm{thr}^\mathrm{NI}$ over the expected background counts ${B}A_\mathrm{geom}$.
Using the above definition of average effective area %(see eq. \ref{eq:AreaEff}) 
we obtain
\begin{equation}
    {S}_\mathrm{NI} = n\mathrm{_{thr}^{NI}}\frac{\sqrt{{B}A_\mathrm{geom}}}{\bar{A}_\mathrm{eff}} = \frac{n\mathrm{_{thr}^{NI}}}{\tau \overline{\varepsilon\eta}_O+(1-\tau)\overline{\varepsilon\eta}_C}\sqrt{\frac{{B}}{A_\mathrm{geom}}}
    \label{eq:NonImagSens}
\end{equation}

\acknowledgments % equivalent to \section*{ACKNOWLEDGMENTS}       
 
%This unnumbered section is used to identify those who have aided the authors in understanding or accomplishing the work presented and to acknowledge sources of funding.  
  
 We acknowledge the financial support of the Italian Space Agency (ASI) and of the Italian National Institute of Astrophysics  (INAF) through the ASI-INAF Agreement n. 2024-17-HH.0. 

% References
%\bibliography{report} % bibliography data in report.bib
\bibliography{xgis} % bibliography data in report.bib
\bibliographystyle{spiebib} % makes bibtex use spiebib.bst

\end{document}